
\def\etal{{\it et al.}\hskip 1.5pt}

\font\machm=cmsy10 \skewchar\machm='60

\def\refset{\parindent=0pt\hangafter=1\hangindent=1em}
\magnification=1200
\hsize=5.80truein
\hoffset=1.20truecm
\newcount\eqtno
\eqtno = 1
\parskip 3pt plus 1pt minus .5pt
\baselineskip 21.5pt plus .1pt
%
%
%
%
\centerline{ \  }
\vskip 0.35in
\centerline{\bf THE CORRELATION FUNCTION OF FLUX-LIMITED X-RAY CLUSTERS}
\vskip 1.00in
\centerline{Neta A. Bahcall and Renyue Cen}
\centerline{\it Princeton University Observatory}
\centerline{\it Princeton, NJ 08544 USA}
\vskip 2.0in
\centerline{To appear in}
\vskip 0.3in
\centerline{\it The Astrophysical Journal Letters, May 1, 1994}
\vskip 0.7in
\vfill\eject

\centerline{\bf ABSTRACT}
	We show that the spatial correlation function of a flux-limited sample
of X-ray selected clusters of galaxies will exhibit a correlation scale that
is smaller than the correlation scale of a volume-limited, richness-limited
sample of comparable apparent spatial density. The flux-limited sample contains
clusters of different richnesses at different distances:  poor groups are found
nearby and rich clusters at greater distances.  Since the cluster correlation
strength is known to increase with richness, the flux-limited sample averages
over the correlations of poor and rich clusters. On the other hand,
a volume-limited, richness-limited sample has a minimum richness threshold,
and a constant mixture of richnesses with redshift. Using the observed
correlation scale for rich ($R\ge 1$) clusters,
$r_o (R\ge 1) = 21\pm 2 h^{-1}$Mpc that was determined from previous
volume-limited studies, we derive for the ROSAT flux-limited X-ray cluster
sample $r_o({\rm flux}{\rm -}{\rm limited}) \approx 14h^{-1}$Mpc,
in agreement with the recently observed value of $13.7\pm 2.3 h^{-1}$Mpc.

\vskip 0.7in
\noindent
Cosmology: large-scale structure of Universe
-- galaxies: clusters of
-- X-rays: galaxies
\vfill\eject

\medskip
\centerline{1. INTRODUCTION}
\medskip
	The correlation function of clusters of galaxies
constrains cosmological models for
the formation and evolution of structure in the universe
(White et al. 1987; Bahcall 1988;
Suginohara \& Suto 1991;
Bahcall \& Cen 1992;
Croft \& Efstathiou 1993).
The cluster correlation function is known to be stronger than
the correlation function of galaxies;
the correlation scale of the rich and rare clusters of richness class
$R\ge 1$ is
$r_o(R\ge 1) = 21\pm 2h^{-1}$Mpc
(Bahcall \& Soneria 1983; Postman et al. 1992; Peacock \& West 1992),
while the galaxy correlation scale is
$r_o(g) = 5h^{-1}Mpc$
(Groth \& Peebles 1977).
The cluster correlation strength is observed
to increase with cluster richness:
rich, rare clusters exhibit stronger correlations
than the more numerous poor clusters and
groups (Bahcall \& Soneira 1983;
Bahcall \& West 1992).
All samples of
clusters studied so far, from poor groups of galaxies to the
richest $R\ge 1$ and $R\ge 2$ Abell
clusters, including the intermediate
richness clusters observed by the new automated cluster
surveys of the EDCC (Nichol et al. 1992) and APM (Dalton et al. 1992)
are consistent with
a universal richness-dependent correlation (Bahcall \& West 1992).
Large-scale cosmological N-body
simulations of galaxy clusters also show the same
dependence of the cluster correlation
function on richness (Bahcall \& Cen 1992; Croft \& Efstathiou 1993).
This richness-dependent correlation function applies
to complete richness-limited samples of clusters
(i.e., clusters above a given richness threshold);
it represents the underlying spatial
distribution of a system of clusters of a given richness class.

	Recently, the spatial correlation function of
a sample of X-ray clusters of galaxies
detected in a flux-limited survey of the ROSAT X-ray satellite
was reported (Romer et al. 1993).
The sample contains all X-ray sources
above a given
X-ray flux threshold
that are
associated with a local
galaxy density enhancement.
A correlation length of
$r_o = 13.7\pm 2.3 h^{-1}$Mpc
was determined for the sample.
Romer et al.(1993) conclude,
from a direct comparison of this correlation length
and the larger length of the rich $R\ge 1$ clusters,
that inconsistencies exist between the two
results, and that the $R\ge 1$ cluster correlation scale
has been overestimated.

	In the present letter we show that
a {\it flux-limited} sample, such as the X-ray sample described
above, differs significantly from a richness-limited sample
(such as the $R\ge 1$ clusters from which the
$21h^{-1}$Mpc scale length was obtained).
The flux-limited sample is a "richness-mixed" sample;
it contains, by definition, poor clusters
nearby and rich clusters at greater distances.
Since cluster correlations depend on richness,
a simple comparison between the correlation
properties of a flux-limited and a richness-limited sample
is inappropriate.
Here we
show that, based on the richness-dependent
correlation function and
a correlation scale of
$r_o (R\ge 1) = 21\pm 2h^{-1}$Mpc
for $R\ge 1$ clusters,
the expected
correlation scale for the above flux-limited X-ray sample is
$r_o ({\rm flux}{\rm -}{\rm limited}) \approx 2/3 r_o (R\ge 1)=14h^{-1}$Mpc,
as is indeed observed.
This first X-ray selected ROSAT cluster sample thus
confirms the richness-dependent cluster correlation function.

\medskip
\centerline{2. THE CORRELATION FUNCTION OF A FLUX-LIMITED}
\centerline{X-RAY SAMPLE}
\medskip
	The richness-dependent
	cluster correlation function
is represented by (Bahcall \& Soneira 1983; Bahcall \& West 1992)
$$\eqalignno{\xi_{cc}(r) &\approx 4N r^{-1.8}\quad ,\quad &(\the\eqtno )\cr}$$
\advance\eqtno by 1
where $\xi_{cc} = Ar^{-1.8} = (r/r_o)^{-1.8}$
is the standard form of the correlation function
(with an amplitude $A$ and a correlation scale $r_o$),
and N is the {\it median} richness of the cluster sample.
Relation (1) applies to complete volume-limited,
richness-limited samples (all clusters
above a threshold richness limit);
it represents the spatial distribution of
clusters of a given richness class.
Similarly, the cluster correlation amplitude
also depends on the mean separation of
clusters, $d$ (where $d = n^{-1/3}$,
and $n$ is the space-density of the cluster sample).
This led to
the universal dimensionless cluster correlation function
(Szalay \& Schramm 1985;
Bahcall \& West 1992)
$$\eqalignno{\xi_{cc}(d) &=0.2(r/d)^{-1.8}=(r/0.4d)^{-1.8}~,~~~~{\rm i.e.,}~~~~
\quad r_o \approx 0.4d~~~~\quad\quad . \quad &(\the\eqtno)\cr}$$
\advance\eqtno by 1

The above-described universal correlation function is
seen both in observations and in model
simulations (Bahcall \& Cen 1992).
It applies to complete volume-limited, richness-limited
samples, where $n$ and $d$ represent the underlying
density and mean separation of a
{\it complete} system of clusters above a given richness threshold.
All the principal cluster samples
analyzed to-date
for which relations (1) and (2) apply are volume and richness
limited
(e.g., Abell, Zwicky, APM, EDCC clusters, and groups;
see summary in Bahcall \& West 1992).

	Recently, a flux-limited correlation function of X-ray clusters was
determined by Romer et al. (1993).
The sample includes
all X-ray sources above a
flux threshold of
$F_x\ge 10^{-12}$ergs~cm$^{-2}$~s$^{-1}$.
A total of 161 sources associated
with some enhancement in the galaxy density distribution are
detected in a 3100 deg$^2$ region;
128 of these systems have measured redshifts.
Romer et al. find
$r_o = 13.7\pm 2.3h^{-1}$Mpc
for the correlation function of this flux-limited redshift sample
(for a correlation slope of
$-1.9\pm 0.4$).
They contrast this correlation scale with the value of
$r_o = 21\pm 2h^{-1}$Mpc
observed for the rich $R\ge 1$ clusters.

	However, the volume-limited and the flux-limited samples
are not expected to yield the same correlation scales,
since the criteria for inclusion in the two samples
are different.
The volume-limited sample includes all clusters
above a given richness threshold; the flux-limited sample
includes a mixture
of richnesses that is a function of redshift.
The {\it average} observed cluster density, $n$,
and mean separation, $d$, of the flux-limited sample
are {\it not} representative of a given richness
system, and thus can not be applied in relations (1) and (2).

	What would the correlation function of such a flux-limited
sample be if the underlying cluster correlation is
represented by the richness-dependent universal correlation
function (relations 1 and 2)?  We address this question below.

	An observational relation exists, as expected theoretically,
between the X-ray
luminosity of clusters and cluster temperature, or mass.
Henry \& Arnaud (1991) find
that, for a Hubble constant of
$H_o=100h$km~s$^{-1}$~Mpc$^{-1}$ (where $h=1$ will be used
throughout),
$L_x({\rm Bol}) = 2.5\times 10^{42} T^{2.7\pm 0.4}$,
where $L_x({\rm Bol})$ is the bolometric X-ray
luminosity of the cluster in erg~s$^{-1}$, and $T$
is the intracluster gas temperature in keV.
Edge \& Stewart (1991) find similar
results, with
$L_x({\rm Bol}) = 2.5\times 10^{42} T^{2.5}$
(see also
Henry et al. 1992 and David \etal 1993 for similar relations).
Converting the bolometric luminosity to the
ROSAT observed energy band of 0.1-2.4 keV (for the
typical range of cluster temperatures $T\sim 2-10$ keV),
we find $L_x(0.1-2.4{\rm keV})\approx 2.8\times 10^{42} T^{2\pm
0.4}$ergs~s$^{-1}$.
The virial mass of a cluster
is proportional to the temperature, $T$, or
to the square of peculiar velocity in the
cluster, $\sigma$:
$M \propto T \propto \sigma^2$
(Sarazin 1988; Bahcall \& Cen 1993; Lubin \& Bahcall 1993).
Lubin and Bahcall find, on average,
$\sigma = 400 T^{0.5}$ km~s$^{-1}$ for the
best $\sigma \propto T^{0.5}$ observed relation.
The virial cluster mass within $1.5h^{-1}$Mpc of the cluster
center, assuming an isothermal density profile, is then
$M(<1.5h^{-1}{\rm Mpc})=2\sigma^2 R(1.5)/G \approx 1.1\times 10^{14} T({\rm
keV})$.
Combining the above relations we find
$$\eqalignno{L_x(0.1-2.4{\rm keV}) &\approx 2.2\times 10^{42}
(M/10^{14}M_\odot)^{2\pm 0.4}{\rm ergs}~{\rm s}^{-1}\quad .\quad &(\the\eqtno
)\cr}$$
\advance\eqtno by 1
The above relation is consistent with the theoretically
expected dependence resulting from
the thermal bremsstrahlung origin of the X-ray emission:
$L_x({\rm Bol}) \propto M_{gas}^2 T^{0.5} \propto M^{2.5}$
(for an approximately constant size of the X-ray emitting region,
and $M_{gas} \propto M$).
This yields $L_x(0.1-2.4) \propto M^{2}$,
consistent with eq. (3).

	Using eq. (3) above, the observed X-ray flux of a cluster is
$$\eqalignno{F_x(0.1-2.4{\rm keV})&\approx 2.2\times 10^{42}
(M/10^{14}M_\odot)^{2\pm 0.4}/4\pi d^2 {\rm ergs}~{\rm cm}^{-2}~{\rm
s}^{-1}\quad ,\quad &(\the\eqtno )\cr}$$
\advance\eqtno by 1
 where $d$ is the luminosity-distance of the cluster.
 The flux is therefore proportional,
approximately, to $F \propto (M/d)^2$.
For a flux-limited sample,
the nearby clusters have a low mass (and thus
a low richness) threshold, and the distant clusters
have a richer threshold.
In a flux-limited sample
the number density of clusters
decreases faster for poor clusters than
for rich clusters.
A volume-limited sample, on the other hand,
has a cluster density that remains constant (in comoving coordinates)
for each richness class.

	To calculate
the expected correlation function of a flux-limited X-ray cluster sample,
we use N-body simulations
that match the correlation function of the observed
richness threshold cluster samples
($R\ge 1$ clusters, EDCC clusters, and APM clusters).
We use a large-scale Particle-Mesh code with a box size of $400h^{-1}$Mpc
to simulate the
evolution of the dark matter.
The box contains $500^3$ cells and $250^3 = 10^{7.2}$ dark matter
particles.  The spatial resolution is $0.8h^{-1}$Mpc.
Details of the simulations are discussed in
Cen (1992) and Bahcall \& Cen (1992).
A model that reproduces the observed mass-function of clusters,
as well as the observed correlation function of the richness-threshold
$R\ge 1$, APM, and EDCC clusters is a low-density,
unbiased CDM model (Bahcall \& Cen 1992).
This model, with $\Omega=0.2$, $h=0.5$, and no bias ($b=1$)
produces the observed
richness-dependent universal cluster correlation
function (eq. 1-2), and is consistent with
other cluster properties such as their mass-function.

	Clusters are selected in the simulation box using
an adaptive linkage algorithm
following the procedure described in Suto, Cen \& Ostriker (1992)
and Bahcall \& Cen (1992);
the cluster mass within
$1.5h^{-1}$Mpc is determined.
In order to extend the volume-limited sample of the underlying
cluster distribution to distances of the most distant X-ray
clusters observed ($z\le 0.25$), a mosaic of
eight $400h^{-1}$Mpc simulation boxes are used, corresponding
to $800h^{-1}$ Mpc on a side.
Within this larger mosaic box, $\sim 3000$ $R\ge 1$
clusters are identified in the simulation (with
$n=6\times 10^{-6}~h^3$Mpc$^{-3}$).

	Each cluster of mass $M$ (within $1.5h^{-1}$Mpc)
is assigned an X-ray luminosity as
given by relation (3), and an X-ray flux---to an observer
at the corner of the box---as given
by (4).  The flux threshold of the Romer et al. (1993) sample
is then applied.
All clusters in the $800h^{-1}$Mpc simulation box
with X-ray flux above this threshold are identified;
they correspond to a flux-limited X-ray
sample similar to the one observed.

	The redshift distribution of the X-ray clusters
in both the observed
and the simulated flux-limited samples
are presented in Figure 1.
Two simulated cases are shown:  one corresponds to relation (4)
(with a 15\% lower amplitude
in the $F_x-M^2$ relation
in order to match the observed {\it average} co-moving
density of X-ray clusters in the range $cz=5000$ to $50000$~km~s$^{-1}$,
$n \sim  5\times 10^{-6} h^3$Mpc$^{-3}$);
the other corresponds to a somewhat shallower slope
than given in relation (4),
$F_x \propto M^{1.7}$, with an
amplitude that yields an average comoving density
of $\sim 8\times 10^{-6}h^3Mpc^{-3}$.
Both simulations yield results that are consistent with the
redshift distribution of the observed flux-limited sample of clusters.
Furthermore, we find that plausible
variations in relation (4) do not produce significant changes in the results.

The {\it mean} richness of the X-ray clusters as a function
of redshift is presented in Figure (2).
In the flux-limited simulations, there is a strong
increase of richness with redshift.
By constrast, the
mean richness in volume-limited samples
is constant.
Approximately 30\% of the clusters are poor ($R\le 0$),
consistent with the observed sample of Romer \etal (1993).

	The correlation function of the simulated
flux-limited sample is presented in Figure 3.
It is compared with the X-ray cluster observations of Romer et al. (1993).
The agreement between the observations and
simulations is excellent.
The simulations yield
$r_o\approx 14h^{-1}$Mpc for the flux-limited sample,
consistent with the observed
$r_o=13.7\pm 2.3h^{-1}$Mpc.
The correlation function of the simulated volume-limited rich $R\ge 1$
clusters (with
$n = 6\times 10^{-6}h^3$Mpc$^{-3}$,
$d = 55h^{-1}$Mpc) is also shown, for comparison;
this function matches well the observed
$R\ge 1$ correlations, with
$r_o \approx 21 h^{-1}$Mpc
(Bahcall \& Soneira 1983; Postman et al. 1992; Peacock \& West 1992).

	Figure 3 shows that the flux-limited
sample exhibits a lower correlation
amplitude than the volume-limited, richness-limited
sample of comparable average number
density.  This is expected due to the ``richness-mixed"
nature of the flux-limited sample.
In the present case, the correlation scales of
the flux-limited and the richness-limited samples satisfy
$$\eqalignno{r_o ({\rm X{\rm -}ray~flux{\rm -}limited}) &\approx {2\over 3} r_o
(R\ge 1)\quad ,\quad &(\the\eqtno )\cr}$$
\advance\eqtno by 1
with
$r_o (R\ge 1) \approx 21 h^{-1}$Mpc
and $r_o({\rm X}{\rm -}{\rm ray}~~{\rm flux}{\rm -}{\rm limited}) \approx
14h^{-1}$Mpc,
as observed.
The results are insensitive to reasonable variations in the
$L_x(M_{\rm cluster})$ relation.
As an example,
we present in Fig. 3
the correlation function for the case
$L_x (0.1-2.4) = 3.5\times 10^{42}(M/10^{14}M\odot)^{1.7}$ergs~s$^{-1}$,
which also matches the observed cluster density distribution (Fig. 1).

\medskip
\centerline{3. CONCLUSIONS}
\medskip
	We show that the spatial correlation function of a
flux-limited sample of X-ray selected clusters
of galaxies exhibits a correlation scale that is smaller than
the correlation scale of a volume-limited, richness-limited sample
of comparable average number density (Fig. 3).
This is expected due to the ``richness-mix" nature
of the flux-limited sample, which contains poor clusters nearby
and rich clusters farther away.

	We show that the correlation scale of the new flux-limited
X-ray cluster sample from ROSAT (Romer \etal 1993)
is expected to be
$r_o ({\rm X{\rm -}ray~~flux{\rm -}limited}) \approx {2\over 3} r_o (R\ge
1)\approx 14h^{-1}$~Mpc, as observed.
We conclude that the new
observations of X-ray clusters from ROSAT
exhibit a correlation function that is consistent with, and is actually
predicted from, the richness-dependent cluster
correlation and $r_o (R\ge 1) \approx 21h^{-1}$Mpc.

It is a pleasure to acknowledge NCSA for allowing
us to use their Convex-240 supercomputer,
on which some of our computations were performed.
R.Y.C is happy to acknowledge
support from NASA grant NAGW-2448 and NSF grant AST91-08103
and N.A.B.  acknowledges support from NSF
grant AST93-15368.

\vfill\eject
\centerline{FIGURE CAPTIONS}
\medskip

\item{Fig. 1--}
Redshift distribution of the flux-limited X-ray clusters.
The X-ray cluster observations of Romer \etal (1993) are shown by
the faint histogram.
Two simulated samples (\S 2) are represented by
the dark and dashed histograms.
$dN(z)$ represents the relative number of clusters
(i.e., fraction of total) in each redshift bin.
(A volume-limited sample yields $dN(z)\propto z^2$).

\item{Fig. 2--}
The {\it mean} richness class of the flux-limited X-ray cluster samples
as a function of redshift.
The two simulated samples (Fig. 1, \S 2)
are shown.
Also presented, for comparison,
are the horizontal lines for the volume-limited,
richness-limited $R\ge 1$ and $R\ge 0$ clusters.

\item{Fig. 3--}
Cluster correlation function: the expected difference
between the cluster correlation function of a volume-limited $R\ge 1$
cluster sample (faint line, with $r_o\approx 21h^{-1}$Mpc),
and the relevant flux-limited X-ray cluster correlations
($F_x\ge 10^{-12}$ergs~cm$^{-2}$~sec$^{-1}$;
dark and dashed lines, $r_o\approx 14h^{-1}$Mpc).
The flux-limited sample
is expected to exhibit weaker correlations than
the $R\ge 1$ complete sample.
The observed flux-limited X-ray cluster
correlations (Romer \etal 1993), shown by the solid dots,
are consistent with the expected flux-limited correlation function
and $r_o(R\ge 1)\approx 21h^{-1}$Mpc.
\vfill\eject

\centerline{REFERENCES}
\refset
Bahcall, N.A., \& Soneira, R.M. 1983, ApJ, 270, 20
\smallskip
\refset
Bahcall, N.A. 1988, ARAA, 26, 631
\smallskip
\refset
Bahcall, N.A., \& Cen, R.Y. 1992, ApJ(Letters), 398, L81
\smallskip
\refset
Bahcall, N.A., \& Cen, R.Y. 1993, ApJ(Letters), 407, L49
\smallskip
\refset
Bahcall, N.A., \& West, M. 1992, ApJ, 392, 419
\smallskip
\refset
Cen, R.Y. 1992, ApJS, 78, 341
\smallskip
\refset
Croft, R.A.C., \& Efstathiou, G. 1993, preprint
\smallskip
\refset
Dalton, G.B., Efstathiou, G., Maddox, S.J., \& Sutherland, W. 1992,
ApJ(Letters), 390, L1
\smallskip
\refset
David, L.P., Slyz, A., Jones, C., Forman, W.
Vrtilik, S.D., \&  Armand, K.A. 1993, ApJ, 412, 479
\smallskip
\refset
Edge, A., \& Stewart, G.C. 1991, MNRAS, 252, 428
\smallskip
\refset
Groth E., \& Peebles, P.J.E. 1977, ApJ, 217, 385
\smallskip
\refset
Henry, J.P., \& Arnaud, K.A. 1991, ApJ, 372, 410
\smallskip
\refset
Henry, J.P., Gioia, I.M., Maccacaro, T., Morris, S.L.,
Stocke, J.T., \& Wolter, A. 1992, ApJ, 386, 408
\smallskip
\refset
Lubin, L.M., \& Bahcall, N.A. 1993, ApJ(Letters), 415, L17
\smallskip
\refset
Nichol, R.C., Collins, C.A., Guzzo, L., \& Lumsden, S.L. 1992, MNRAS, 255, 21
\smallskip
\refset
Peacock, J., \& West, M. 1992, MNRAS, 259, 494
\smallskip
\refset
Postman, M., Huchra, J., \& Geller, M. 1992, ApJ, 384, 404
\smallskip
\refset
Romer, A.K., Collins, C.A., Boringer, H., Ebeling, H.,
Voges, W., Cruddace, R.G., \& MacGillivray, H.T. 1993, preprint
\smallskip
\refset
Sarazin, C.L. 1988, in ``X-ray Emission in Clusters of Galaxies"
(Cambridge: Cambridge University Press)
\smallskip
\refset
Suginohara, T., \& Suto, Y. 1992, PASJ, 43, L17
\smallskip
\refset
Suto, Y., Cen, R.Y., \& Ostriker, J.P. 1992, ApJ, 395, 1
\smallskip
\refset
Szalay, A., \& Schramm, D.N., Nature, 314, 718
\smallskip
\refset
White, S.D.M., Frenk, C.S., Davis, M., \& Efstathiou, G. 1987,
  ApJ, 313, 505
\smallskip
\refset
\vfill\eject\end